# Integrated LLM-Based Intrusion Detection with Secure Slicing xApp for Securing O-RAN-Enabled Wireless Network Deployments

Joshua Moore, Aly Sabri Abdalla, Prabesh Khanal, and Vuk Marojevic
Dept. of Electrical and Computer Engineering, Mississippi State University, USA
Emails: {jjm702; asa298; pk571; vuk.marojevic}@msstate.edu

*Abstract*—The Open Radio Access Network (O-RAN) architecture is reshaping telecommunications by promoting openness, flexibility, and intelligent closed-loop optimization. By decoupling hardware and software and enabling multi-vendor deployments, O-RAN reduces costs, enhances performance, and allows rapid adaptation to new technologies. A key innovation is intelligent network slicing, which partitions networks into isolated slices tailored for specific use cases or quality of service requirements. The RAN Intelligent Controller further optimizes resource allocation, ensuring efficient utilization and improved service quality for user equipment (UEs). However, the modular and dynamic nature of O-RAN expands the threat surface, necessitating advanced security measures to maintain network integrity, confidentiality, and availability. Intrusion detection systems have become essential for identifying and mitigating attacks. This research explores using large language models (LLMs) to generate security recommendations based on the temporal traffic patterns of connected UEs. The paper introduces an LLM-driven intrusion detection framework and demonstrates its efficacy through experimental deployments, comparing non-fine-tuned and fine-tuned models for task-specific accuracy.

*Index Terms*—Intrusion detection, LLM, latency, Open Artificial Intelligence Cellular, O-RAN, security, slicing, xApp.

## I. Introduction

The Open Radio Access Network (O-RAN) is transforming the telecommunications landscape by promoting openness, flexibility, and intelligent closed-loop RAN optimization [1]. The main components of O-RAN include the Central Unit (CU), Distributed Unit (DU), and Radio Unit (RU), which together enable the disaggregation of traditional RAN functions for greater flexibility and multi-vendor interoperability. Additionally, the RAN Intelligent Controllers (RICs), the near-real time (RT) RIC and the non-RT RIC, enable artificial intelligence (AI)-driven optimization and management, enhancing network efficiency and adaptability. By decoupling hardware and software components and enabling multi-vendor deployments, O-RAN reduces capital expenditures while enhancing network performance as well as allowing operators to adapt rapidly to new technologies [2].

One such innovation is intelligent network slicing, a technique that allows operators to partition their networks into custom, isolated slices, each tailored to a specific application or quality of service (QoS) requirement. This capability is further enhanced by the ability of the RICs to optimize the resource allocation within network slices, enabling efficient resource utilization, which improves service quality for connected user equipment (UEs).

However as illustrated in Fig. 1, while O-RAN establishes a more dynamic and modular architecture, it also increases the threat surface [3]. For instance, open interfaces in O-RAN enable seamless interoperability among equipment, but this openness also introduces significant security risks, as attackers may exploit these interfaces to inject malicious traffic or intercept sensitive communications. The reliance on over-the-air (OTA) transmissions between network elements exposes O-RAN to jamming, eavesdropping, and replay attacks, among others, which can compromise the integrity and availability of data. Furthermore, the intelligence of RICs presents additional vulnerabilities, as adversaries might manipulate AI models or introduce adversarial inputs, disrupting network optimization and decision-making processes. Advanced security measures must thus be considered to ensure the integrity, confidentiality, and availability of the network. Intrusion detection systems have become a critical component in identifying and mitigating malicious attacks to both traditional RAN and to future O-RAN deployments.

Recently, large language models (LLMs) have received significant attention from academia and industry for enabling intelligent decision-making across a wide array of fields, including education, finance, healthcare, and biology [4]. The capabilities of LLMs to process vast amounts of data and generate insights have positioned them as powerful assets for addressing complex challenges in these domains. In the field of wireless communications, LLMs have demonstrated remarkable potential in managing and optimizing network performance across diverse deployment scenarios [5]. Specifically, LLM-driven wireless network optimizations have been leveraged for edge intelligence [6], network intrusion detection [7], and reconfigurable intelligent surface deployments [8].

This research explores the unfolding possibilities of O-RAN in leveraging the capabilities of LLMs to generate security recommendations based on temporal traffic patterns of connected UEs. The goal is to evaluate the practical application of this emerging technology in enabling secure slicing, supporting heterogeneous traffic types on a shared O-RAN infrastructure. In this paper, we introduce a framework for LLM-driven intrusion detection and, through experimental deployment, we show the efficacy of this approach for task-specific instructions comparing non-fine-tuned and fine-tuned models.

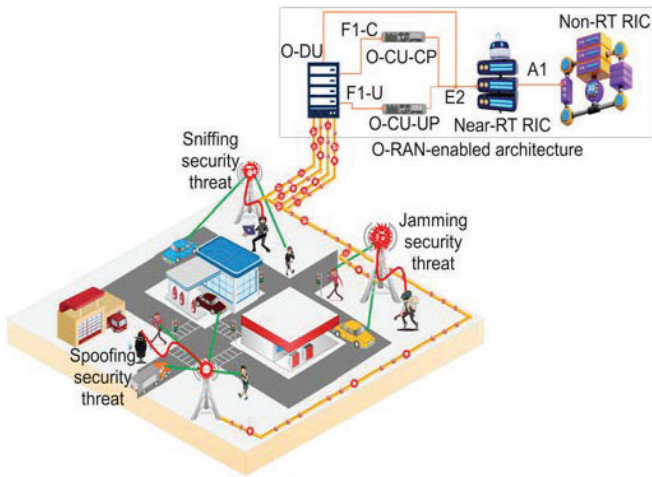

Fig. 1: The O-RAN architecture with wireless attacks.

The main contributions of this paper are as follows:
- We propose a framework for LLM-based intrusion detection in the O-RAN context.
- We introduce an integrated framework composed of three xApps—KPIMON, LLM-based intrusion detection, and secure slicing—deployed in the near-RT RIC for collecting real-time data, detecting malicious activities, and isolating any detected intruder.
- Through an experimental deployment that leverages the Open Artificial Intelligence Cellular (OAIC) platform we demonstrate the effectiveness of this approach in safeguarding legitimate users in the presence of intruders.

The rest of this paper is organized as follows: Section II provides background related to intrusion detection in O-RAN as well as work leveraging LLMs for intrusion detection. In Section III, we introduce the experimental deployment and framework integration. Section IV details results obtained from testing the LLM as well as discusses the major findings and outcomes from testing various pre-trained models and one instruction tuned open source model. Section V presents the concluding remarks and future research directions.

## II. BACKGROUND AND PRIOR WORK

Intrusion detection systems in the O-RAN architecture are designed to process diverse data streams from multiple sources within the network, such as the RU, DU, and CU. These data streams provide a rich set of information, including network traffic patterns, signaling messages, and system logs, which are critical for comprehensive threat monitoring and analysis. Analysis engines employ a mix of techniques, including signature-based detection for known threats, anomaly-based models powered by machine learning to identify unusual behaviors, and heuristic or behavior-based methods to recognize deviations from expected patterns. The combination of these engines ensures robust detection capabilities across a wide range of threat scenarios, from traditional attacks to sophisticated, zero-day exploits. Upon identifying potential intrusions, the intrusion detection systems can initiate various response mechanisms tailored to the severity and nature of the threat. Responses may include automated actions, such as isolating compromised components, triggering alarms for human intervention, or adapting network configurations to mitigate the impact [9].

Large language models offer potential to increase the capabilities of existing intrusion detection mechanisms within O-RAN. A proposed framework [10] uses a pre-trained LLM to automate intrusion detection in 5G networks by selecting relevant features, processing data, building prompts, and extracting decisions. In-context learning, a method where a pre-trained LLM is guided to improve performance on specific tasks by integrating examples directly into the prompts without altering the model's parameters, is employed to improve detection accuracy using labeled examples and task-specific guidance. The work shows the capability of LLMs for feature extraction with a collected dataset but does not actualize the integration into a 5G network and uses pre-trained closed-source models for in-context learning results. Similarly, [11] shows the potential of LLMs in detecting anomalies in 5G network traffic. The work presents a centralized approach as well as a federated learning approach due to the edge device constraints. The results confirm that, while federated learning may have slightly lower accuracy than centralized approaches, it offers enhanced data privacy and scalability.

Llama 2 is used for detecting distributed denial of service (DDoS) attacks and improved using human feedback and compared against several established methods using the CIC-IDS2017 dataset [12]. The results demonstrate that Llama 2 achieves high accuracy and efficiency in detecting DDoS attacks with a performance comparable to long short-term memory, convolutional neural network, and deep neural network models, while also being suitable for real-time network traffic monitoring.

Unlike earlier studies that primarily explore anomaly detection or feature extraction from from previously collected network traffic datasets, this research proposes a comprehensive framework for LLM-driven intrusion detection specifically within the O-RAN context.

## III. O-RAN INTRUSION DETECTION FRAMEWORK

### A. LLM-based Intrusion Detection and Mitigation

To enhance security in O-RAN deployments, we propose a Large Language Model-based Intrusion Detection and Mitigation framework (LLM-ID), which leverages the reasoning capabilities of LLMs for real-time anomaly detection and response. This research introduces a comprehensive framework tailored specifically to the O-RAN context. By deploying this framework on the OAIC platform, we demonstrate its real-world applicability, going beyond theoretical models. Additionally, the research highlights how fine-tuning LLMs can significantly improve the accuracy of task-specific security recommendations for securely sliced RANs. Through experimental deployment, we provide valuable insights into automating the protection of resources for legitimate O-RAN users. The LLM-ID framework seamlessly integrates with the network control infrastructure, enabling adaptive and context-aware intrusion detection, with the LLM-ID xApp playing a central role in the automated intrusion detection system, as shown in Fig. 2.

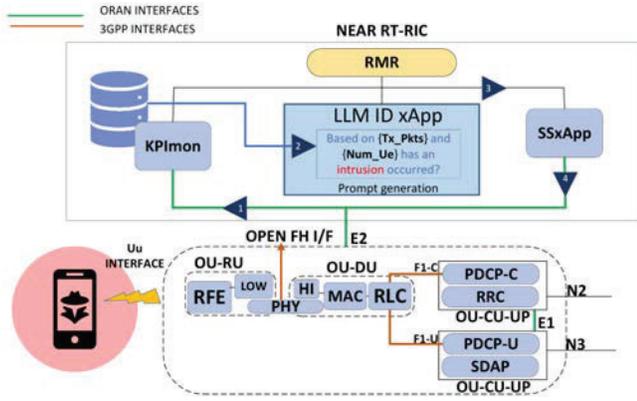

Fig. 2: Integrated LLM-based intrusion detection (ID) with secure slicing deployed in OAIC's near-RT RIC.

The framework consists of three distinct xApps and other major components performing specific roles as follows:

- Database: The database is used to store information collected from KPIMON, which can be used by other xApps.
- RIC Message Router (RMR): Acts as the backbone for message exchanges between various xApps, facilitating data transfer and communication within the near-RT RIC.
- LLM-ID xApp: This xApp analyzes the received data (e.g., TX Pkts and NUM UEs) to determine if an intrusion has occurred. The xApp uses natural language processing to generate security prompts and alerts based on data patterns.
- KPIMON xApp: This xApp collects information from the RAN.
- SSxApp: A security-focused xApp [13] that responds to prompts and alerts generated by the LLM-ID xApp. It can initiate appropriate security measures to mitigate or address potential threats.

Fig. 3 shows the interactions between the components of the OAIC RIC for the detection of intrusions based on collected metrics from the RAN, the mitigation of DDoS attacks over the E2 interface, and the recovery of legitimate UEs operating within a targeted network slice. The KPIMON xApp collects periodic reports from the E2 node over the E2 interface and stores this information in a database. The LLM-ID xApp collects this stored information where it then extracts relevant KPIs to be used in an instruction tuned prompt assessing if an intrusion has occurred based on the number of Tx Pkts that were transferred from the UE to the RAN indicating the number of requests for resources that the UE made as well as the number of active UEs. After being inferenced by the prompt, the model is instructed to generate only a single word outcome, either *Malicious* or *Legitimate*, to cut down the time it takes to get a full response from the model. If the UE is considered legitimate, the next prompt is constructed and the model is inferenced again. If the UE exhibits anomalous traffic behavior, the LLM-ID xApp forwards a request to the secure slicing xApp through the RMR which then generates a control message to unbind the offending UE from the slice and bind it to a slice with no resources, securing the slice and other legitimate UEs from further harm.

### B. Experimental Deployment and Fine-Tuning

To demonstrate the potential of LLM-ID xApp's integration, we extend upon the OAIC platform [14], an openly accessible framework which integrates wireless communication hardware and offers a comprehensive framework for monitoring, analyzing, and securing network slices through xApps deployed in OAIC's near-RT RIC, which is based on the O-RAN Software Community's RIC [15]. The OAIC testbed incorporates the following features:

- A 3GPP-compliant end-to-end 5G-based RAN and core network implementing open-source software protocol stacks srsRAN and Open5GS.
- An O-RAN-compliant control network architecture augmented with OAIC's near-RT RIC to host, maintain, and employ different AI-based RAN intelligent controllers as xApps. The near-RT RIC is implemented with the standard E2 interface and service models (SMs) that ensures seamless communication between the RIC and the E2 nodes by coordinating the flow of key performance measurement (KPM) reports from the RAN and delivery of control actions from xApps.
- Reconfigurable RAN functionalities, such as slicing and scheduling, through modular open Application Programming Interfaces (APIs).

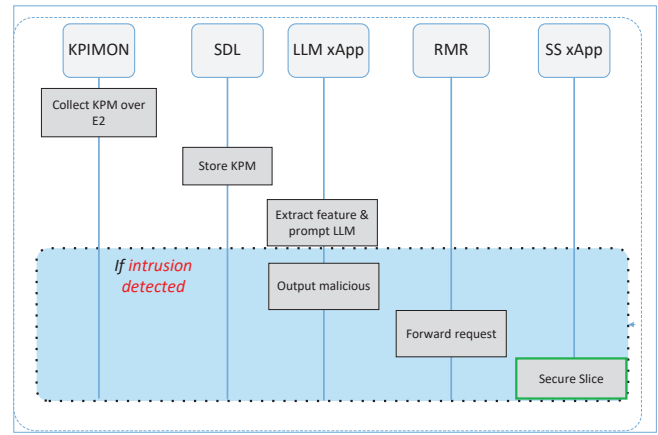

Fig. 3: Process flow of integrated LLM-based ID with secure slicing framework.

We deploy a wireless communication system composed of a single base station with 3 UEs implemented over B210 Universal Software Radio Peripherals (USRP). These UEs are configured within a single slice as Enhanced Mobile Broadband (eMBB) users, which require continuous high-throughput requirements to support video streaming, immersive applications, and other data-intensive services. The bandwidth of the base station is set to 20 MHz with 100 RBs divided equally between the three users of the eMBB slice, where the eMBB user demands a 10 Mbps bit rate. The near-RT RIC is a modified version of the "E" release from the Open Source Community (OSC), enhanced with OSC's Key Performance Indicator Monitoring (KPIMON) xApp for performance monitoring and a custom secure slicing xApp (SSxApp) to enforce security measures and handle resource isolation upon detection of intrusion.

*1) Data Collection:* Data collection involves configuring the near-RT RIC to gather key performance indicators (KPIs), including packet transmission rates, resource allocation metrics, and UE connection status, from the RAN. We implement scripts for automated data logging and used Python tools for further analysis and visualization. We position the attacker at a fixed distance to ensure consistent signal propagation and data variability, monitoring performance metrics across various distances. We move the legitimate UEs with respect to the experimental base station in the laboratory to produce data sets with varying modulation and coding schemes. Collected data is preprocessed with Python scripts to remove obvious errors, and extract key parameters for fine-tuning the LLM.

*2) Fine-Tuning the LLM:* Abiding by the O-RAN specifications [16], we fine-tune and deploy our LLM-based intrusion system offline on a GPU-accelerated server. The server is equipped with an Nvidia RTX 4090 GPU (24 GB VRAM), an AMD Ryzen 7 six-core CPU, and 128 GB of system RAM; it runs Ubuntu 22.04 LTS with a low-latency Linux kernel.

Gemma 2, an open-source LLM, is fine-tuned using the unsloth framework for a specialized instruction-based intrusion detection task in O-RAN. Fine-tuning Gemma 2 involves adapting the pre-trained model to recognize patterns indicative of security threats in the network traffic. This process trains the model on a data set containing labeled instances of both normal and malicious activities, with randomized variables such as the number of user devices and transmitted packets to expose the model to diverse traffic patterns, enhancing generalization and reducing overfitting. During this fine-tuning, model parameters are adjusted to optimize the ability to distinguish between legitimate and abnormal traffic [17].

This fine-tuning leverages standardized O-RAN KPIs, including downlink and uplink bytes (DL BYTES, UL BYTES), physical resource blocks used (DL PRBS, UL PRBS), packet counts (TX PKTS, RX PKTS), transmission and reception errors (TX ERRORS, UL ERRORS), and active user counts (NUM UEs). These metrics enhance the model's understanding of network behavior. Aligning Gemma 2's learning objectives with KPIs, such as TX PKTS and NUM UEs, collected from comprehensive OTA testing, optimizes the model's effectiveness in identifying intrusions and generating security recommendations. The use of standardized KPIs ensures that the fine-tuning aligns with the O-RAN philosophy and is relevant to real-world deployments.

## IV. EXPERIMENTAL DEPLOYMENT AND RESULTS

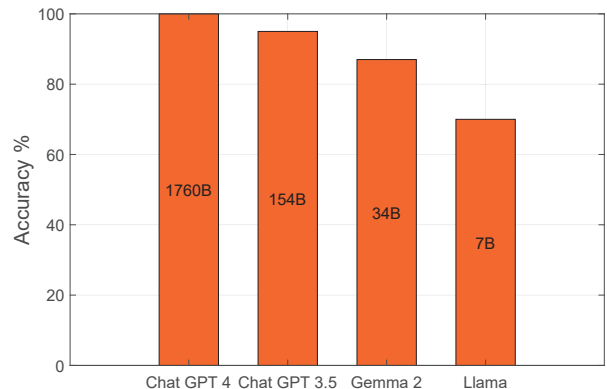

(a) accuracy of base models with "few shot prompting"

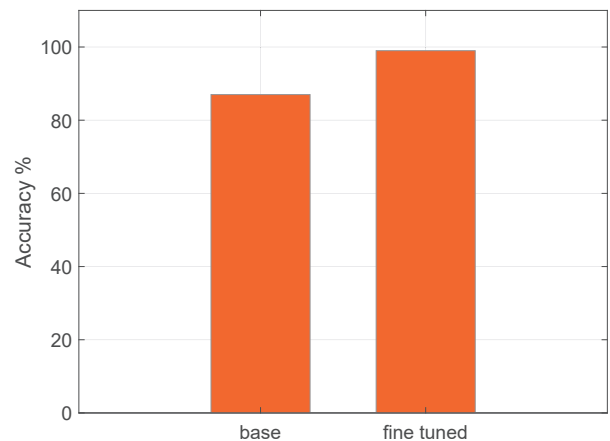

(b) accuracy after fine tuning

Fig. 4: Comparison of model accuracy with few-shot prompting (a) and fine-tuning (b) for detecting anomalous user traffic flows.

We compare easily accessible closed and open source LLMs in two scenarios: (1) Non-fine-tuned models with few shot prompting techniques and (2) a fine-tuned open source model, Gemma 2, tuned for a task-specific prompt. We assess this fine-tuned model on the OAIC platform, where we deploy the LLM-ID and SSxApp and contrast it with a static threshold for intrusion detection and secure slicing.

The experimental setup involves a scenario where a malicious user device (UE) generates a flood of network requests to simulate a Denial of Service (DoS) attack. In this setup, the malicious UE overwhelms the resources allocated to specific network slices by sending a high volume of requests. This results in degraded service quality for legitimate users, demonstrating an effective strategy for disrupting service as shown in prior research [18]. The attack exploits the absence of robust authentication mechanisms within the resource allocation policies of the slicing xApp. This vulnerability allows the attacker to disrupt the fair distribution of resources that the xApp is designed to ensure for connected user devices, ultimately causing potential service disruptions for legitimate users.

As shown in Fig. 4a, LLMs can perform well when instructed with a prompt that includes a clear example, such as defining the boundary of legitimate UEs in a network slice. This provides the model with context and helps it understand the task-specific constraints, leading to more accurate and relevant outputs. Specifically, we give the model an example of a legitimate UE based on its TxPkts and one malicious so that the model has the context of the boundary for TxPkts which is a common KPI in O-RAN.

The given prompt is:

> PLEASE ONLY OUTPUT IN A WORD with TX Pack limits of 312 for 1 UE and 624 for 2 UEs, check if the following {NumUE} and {TXPackets} meet these bounds. If within bounds output *Legitimate* (input $\leq$ bounds) or *Malicious* (input $\geq$ bounds if exceeded).

This type of prompting is called "few-shot prompting" or "demonstration-based prompting."

Larger models, such as Chat GPT 4 (1760B parameters) achieve higher accuracy (100%) in intrusion detection, demonstrating the advantage of complex architectures in capturing intricate patterns (Fig. 4a). Chat GPT 3.5 (154B) follows with 95% accuracy, showing strong performance but lagging behind Chat GPT 4. Gemma 2 (34B) achieves 87%, indicating that smaller models can still perform well, though less effectively than the larger ones. Llama (7B) performs the weakest at 70% accuracy, reflecting the limitations of smaller models in such complex tasks. Open-source models Gemma 2 and Llama provide flexibility and accessibility, while closed-source models Chat GPT 4 and Chat GPT 3.5 offer higher performance but are less customizable and accessible for research and development.

As illustrated in Fig. 4b, instruction tuning an open-source model can significantly enhance task-specific accuracy. Using demonstration-based prompting, the Gemma 2 model initially achieves an accuracy of 87% without any prior training. However, after fine-tuning, the accuracy for the same task increases to 99%, showcasing the effectiveness of the fine-tuning process.

Fig. 5 shows three UEs operating within the same slice with a maximum downlink throughput of 10 Mbps. Around 170 seconds into the experiment, one of the UEs becomes greedy and starts to consume more resources than prescribed, stealing resources from the other UEs operating within the same slice. Based on the traffic metrics collected from this UE, the LLM is able to identify the abnormality in the Tx Pkts as well as the throughput and classify the UE as an intruder. This is then used for the secure slicing xApp to mitigate further damage from occurring within the slice due to its ability to move the malicious UE to a slice with no resources. After that the malicious UE is eliminated, the legitimate UEs are able to quickly recover and access the previously captured resources by the malicious user. The average detection and response time is 239.66 ms, which reflects the time from the receipt of KPM reports to the mitigation of the intrusion. The KPImon xApp collects the KPM reports, which are generated at intervals between 1 ms and 1 second, and then the LLM-ID xApp analyzes these reports to detect potential intrusions. Upon detecting an anomaly, the SSxApp isolates the malicious UE by reallocating resources, effectively slicing it away from the network and minimizing disruption to legitimate users.

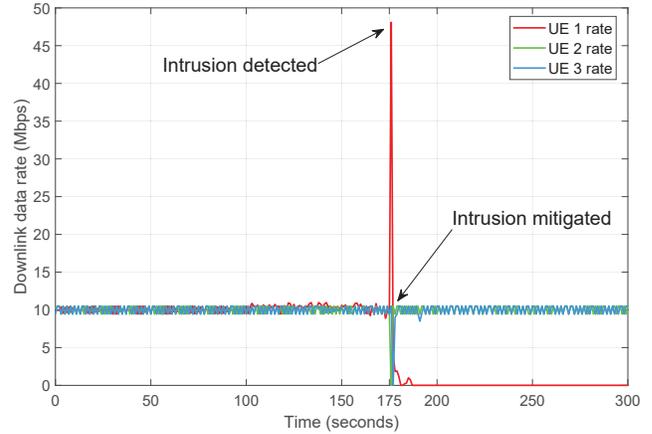

Fig. 5: Achieved data rate performance over time for legitimate UEs (UE2 and UE3) and malicious UE (UE1), illustrating the LLM-based ID and SSxApp mitigation events.

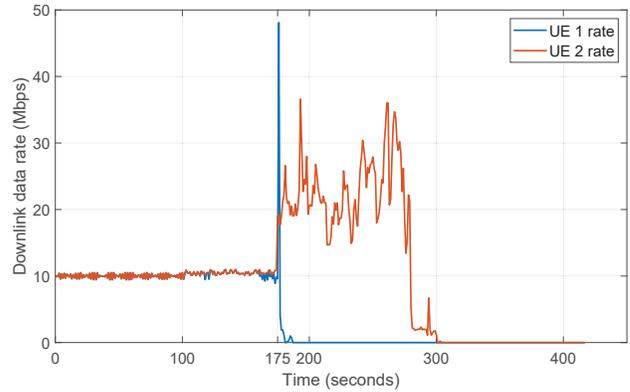

Fig. 6: Achieved datarate performance over experiment time for the LLM-based ID (UE 1) compared to a static ID method (UE 2).

Fig. 6 compares the LLM-enabled intrusion detection with secure slicing (UE1) against intrusion detection based on a static threshold (UE2). This comparison of throughput over simulation time shows the difference in time to detect an intrusion and to move the UE to a slice with no resources. The LLM is the fine-tuned Gemma 2 model tuned for the task of identifying anomalous UE traffic flows. The 10 Mbps intended downlink throughput for the connected UE is initially observed. Anomalous behavior starts around 175 s for each separate trial. UE1 is analyzed by the proposed LLM-ID system, which identifies it promptly as malicious to initiate the process or rebinding it to a zero-resource slice. UE2, which is processed by the intrusion detection based on the static threshold, is not marked malicious until the 280 s mark. This longer identification is due to false positives, which we have shown in our previous work [19] in which the static threshold was used. Multiple KPM reports can be collected to mitigate these false positives, leading to a slower detection. The LLM-enabled intrusion detection system, on the other hand, can accurately asses if an intrusion has occurred with only one KPM report. This capability is driven by the model's advanced contextual understanding, which enhances its ability to distinguish between normal network fluctuations and true security threats with greater precision, minimizing the reliance on multiple data points.

By leveraging the LLM's ability to analyze network traffic and identify malicious patterns in real-time, the system helps ensure compliance with service level agreements across network slices. The proactive security approach not only mitigates the risk of attacks but also optimizes detection time and network performance, ensuring integrity within network slices. This approach underscores the critical role of intelligent, adaptive network management for safeguarding O-RAN deployments.

## V. Conclusions

O-RAN represents a transformative advancement in telecommunications by promoting openness, flexibility, and intelligent resource management to support multi-vendor deployments, dynamic network slicing, and network intelligence. While these innovations significantly enhance network performance and adaptability, they also introduce new security challenges due to the system's modular and dynamic nature. This paper has presented a novel approach of using LLMs to strengthen O-RAN intrusion detection capabilities, effectively analyzing and generating security recommendations based on temporal traffic patterns of UEs. An intelligent framework composed of the proposed LLM-based intrusion detection xApp, the KPM report collection xApp, and the secure slicing xApp has been presented to effectively detect anomalies and malicious behavior and then isolate intruders to secure the network resources. Experimental results over the OAIC testbed have demonstrated the improved accuracy of an open-source model that is fine-tuned for the specific task of identifying anomalous traffic patterns, underscoring the potential of LLM-driven frameworks in enhancing the security and resilience of future O-RAN deployments.

## Acknowledgement

This work was supported in part by NSF award 2120442 as well as NSF and Office of the Under Secretary of Defense (OUSD) – Research and Engineering, under Grant ITE2326898, as part of the NSF Convergence Accelerator Track G: Securely Operating Through 5G Infrastructure Program.